\documentclass[preprint,prb]{revtex4}
\usepackage{bm}
\begin{document}
\title{On the structure of the energy distribution function in the hopping 
regime}
\author{O. Bleibaum}
\affiliation{Institut f\"ur Theoretische Physik, 
Otto-von-Guericke Universit\"at
Magdeburg, PF 4120, 39016 Magdeburg, Germany}
\author{H. B\"ottger}
\affiliation{Institut f\"ur Theoretische Physik, 
Otto-von-Guericke Universit\"at
Magdeburg, PF 4120, 39016 Magdeburg, Germany}
\author{V. V. Bryksin}
\affiliation{A. F. Ioffe Physico-Technical Institute, St. Petersburg,
Politheknicheskaya 26, 194021 St. Petersburg, Russia}
\begin{abstract}
The impact of the dispersion of the transport coefficients on the structure
of the energy distribution function for charge carriers far from equilibrium
has been investigated in effective
medium approximation for model density of states. The investigations shows 
that two regimes can be observed in energy relaxation processes. Below a 
characteristic temperature the structure of the energy distribution function 
is determined by the dispersion of the transport coefficients. Thermal energy 
diffusion is irrelevant in this regime. Above the characteristic temperature
the structure of the energy distribution function is determined by energy
diffusion. The characteristic temperature depends on the degree of disorder
and increases with increasing disorder. Explicit expressions for the
energy distribution function in both regimes are derived for a constant
and an exponential density of states. 
\end{abstract}
\maketitle
%%%%%%%%%%%%%%%%%%%%%%%%%%%%%%%%%%%%%%%%%%%%%%%%%%%%%%%%%%%%%%%%%%%%%%%%%%
\section{Introduction}
A characteristic feature of many disordered materials is the strong dependence
of their transport properties on the frequency of the applied external 
electric field for low frequencies. In that disordered materials differ from 
ordered materials, which typically only show a  pronounced frequency 
dependence at high frequencies. Also the character of the impact of an
alternating electric field on the transport coefficients in disordered 
materials is different from that in ordered ones. Whereas in ordered materials
an increase of the frequency always reflects in a decrease of the 
conductivity  in disordered materials an increase of the frequency 
leads to  an increase of the real part of the conductivity. The sensibility
of the transport coefficients to a change of the  frequency increases with 
increasing amount of disorder. Therefore, a particular strong frequency 
dependence is observed in the strongly localized regime, in which transport
proceeds by hopping between localized states, such as impurity bands, 
Anderson insulators, glasses, and polymers
(see, e.g, Refs.[\onlinecite{Boettger,Pollak,Brom,Porto,Dyre1}]).

The same frequency dependence, which governs the response to an alternating
electric field, determines also the relaxation properties of the system.
Accordingly, the dispersion of the dynamical conductivity in the hopping regime
reflects itself also in relaxation experiments close to equilibrium. It 
manifests itself, e.g., in transient current 
experiments\cite{Boettger,Thomas1}, which test the diffusion properties of 
particle packets in a random environment. In such experiments the dispersion 
of the transport coefficients leads to dispersive
transport, this is to deviations from the conventional 
relationship $\langle R^2(t)\rangle\propto t$ for $t<t_{perc}$, where 
$\langle R^2\rangle$ is the mean squared deviation of a particle packet at 
time $t$, which was a 
delta-pulse at time $t=0$, and $t_{perc}$ is the 
percolation time\cite{Pollak}. Accordingly, the diffusion process breaks up 
into two regimes, a regime of anomalous diffusion, which takes place for 
$t<t_{perc}$, and a regime of normal diffusion, which takes place for 
$t\gg t_{perc}$. Since the percolation time depends exponentially on the 
disorder parameter it depends strongly on the material in question, i.e.
there is  no typical time scale for it in the experiments. In practice time 
scales between $10^{-8}$s and 5s have been observed (see discussion of 
experiments on p.216 of Ref.[\onlinecite{Boettger}]).

Far from equilibrium mainly the effective rate analysis of Ref.
[\onlinecite{Monroe}] has been applied to analytical investigations of 
relaxation
phenomena. This method assumes that the relaxation process far from 
equilibrium is not affected by the dispersion of the transport 
coefficients, and that therefore, the transport coefficients to be used 
in modeling the relaxation are the same as those at zero
frequency. We would like to mention that despite this fact also
dispersive transport has been discussed in applications of the effective
rate analysis, but that the nature of the dispersive transport discussed 
there, e.g. in Ref.[\onlinecite{Monroe}], is very different from that
in transient current experiments close to equilibrium. Whereas
the dispersive relaxation close to equilibrium results from
the frequency dependence of the transport coefficients, and thus 
reflects the peculiarity that the charge carriers in hopping systems 
try to optimize their jumps, the dispersion
in the effective rate analysis results solely from the fact that the 
particles move in an inhomogeneous distribution of sites in the 
four dimensional position-energy space. In this case dispersion arises 
since the transport coefficients
depend on the instantaneous position of the particles in energy space, which
changes with time. Therefore, similar dispersive contributions
are also obtained for the relaxation of charge carriers in ordered 
systems. Clearly, they are much larger in the hopping regime 
than in ordered systems, since the transport coefficients  
depend much stronger on the energy in the hopping regime than in 
ordered systems.

Given that the analytical results of the effective rate analysis correlate
well with the results of numerical experiments for systems with strongly
energy dependent density of states the question arises whether
the dispersion of the transport coefficients is important in  
relaxation processes far from equilibrium or not, and if so, which quantities
are determined by it. To answer this question at zero temperature we have 
derived an equation for the energy  distribution function in the hopping 
regime in Ref.[\onlinecite{BBB2}] and investigated it further in the 
Refs.[\onlinecite{BBB1}] and [\onlinecite{Haba}]. The most important property 
of this equation is this the shape of the energy distribution function
is entirely determined by the dispersion of the transport coefficients.
The dispersion determines the first moment of the energy distribution function
for systems with weakly energy dependent density of states and
the second and the higher moments both for systems with weakly energy dependent
density of states and for systems with strongly energy dependent density of 
states. These moments can be investigated experimentally 
by line width measurements in time resolved photoluminescence spectroscopy
experiments. In an exponential density of states the dispersive nature
of the energy relaxation leads to a soliton like motion of a particle packet 
in energy space at zero temperature. 

It is natural to ask whether these
results apply also to systems at finite temperatures or not. If so, these
results would provide a further characteristic, which would set disordered 
hopping systems apart from other physical systems. The latter generally
share the feature that the structure of the energy distribution function is
determined by energy diffusion processes\cite{Kampen,Landau}. In the 
diffusion approximation,e.g. this fact reflects in that the second derivatives
with respect to the energy are as important as the first derivatives
and therefore  can not be ignored\cite{Landau}. Below we investigate 
this question analytically for charge carriers far from the Fermi energy
in a constant density of 
states and charge carriers in an exponential density of states. For these
model densities of states we show that there is a characteristic temperature,
which sets two regimes apart, a regime where the structure of the energy
distribution function is determined by the dispersion of the transport
coefficients and a regime where the structure of the distribution function
is governed by the energy diffusion current. The characteristic temperature
discriminating these regimes is finite, depends on the amount of disorder, 
and increases with increasing disorder.

%%%%%%%%%%%%%%%%%%%%%%%%%%%%%%%%%%%%%%%%%%%%%%%%%%%%%%%%%%%%%%%%%%%%%%%%%%
\section{The model}
We consider strongly localized non-equilibrium charge carriers far from the
Fermi-energy. Such charge carriers can, e.g., be produced by illumination.
After their excitation the charge carriers relax by phonon-assisted hops.
We assume that the number of charge carriers is small, so that we can
neglect Fermi-correlation. In this case the transport can be described by
the simple rate equation
\begin{equation}\label{M1}
\frac{dn_m}{dt}=\sum_n[n_nW_{nm}-n_mW_{mn}].
\end{equation}
Here $n_m$ is the probability to find a particle on the site $m$ with
site energy $\epsilon_m$ and position vector $\bm{R}_m$,
\begin{equation}\label{M2}
W_{nm}=\theta(\omega-|\epsilon_{nm}|)\nu\exp(-2\alpha|{\bm{R}_{nm}}|
+\frac{\beta}{2}
(\epsilon_{nm}-|\epsilon_{nm}|)),
\end{equation}
$\epsilon_{nm}=\epsilon_n-\epsilon_m$,
$\bm{R}_{nm}=\bm{R}_n-\bm{R}_m$, $\alpha^{-1}$ is the localization
length, $\nu$ is the attempt-to-escape frequency, and $\beta$ is the
inverse temperature. Both the positions of the sites and the site energies
are random quantities. We assume that the sites are distributed randomly
in space and that the site energies are distributed according to a
distribution function, which is proportional to the density of states
$N(\epsilon)$ of the system. 

In writing down Eq.(\ref{M2}) we have also assumed that there is a maximal 
amount of energy $\omega$, which can be transferred from the electron to the 
phonon 
system, or vice versa, in one hop. The introduction of this energy 
scale is based on the observation that the electron phonon coupling constant
tends rapidly to zero for phonons with large momentum $q$. The electron
phonon coupling constant is a function which depends on both the overlap
between the electron wave function and the phonon wave function and on the
the Fourier transformed electromagnetic potential, which provides the coupling
between the electron and the phonon wave. The overlap integral tends to zero
for phonons with wave vector $q>2\alpha$, and thus renders the modes with 
$q>2\alpha$
ineffective. The Fourier transformed potential is usually considered as a
constant, in line with the deformation potential approximation. From the
physical point of view this means that the electrons can move in such a way 
that they can keep the system neutral. In the strongly localized regime,
however, there are no electrons which can move fast enough to 
screen out the alternations of the electric field produced by the phonon 
wave for most $q$. Accordingly, the Fourier-transformed Coulomb potential
is a function which drops rapidly to zero already for tiny $q$. Due to these
facts hops with large energy transfer can happen, but are not characteristic.
In our approximation we therefore ignore such jumps. Since at present
there are no further investigations in the literature on the energy scale
$\omega$ we take the point of view that $\omega$ is a phenomenological 
parameter, which can be determined by an experiment.  

In our investigation we focus on energy relaxation processes.
Quantities of interest are, e.g.,  the mean energy
\begin{equation}\label{M3}
\langle\epsilon\rangle(t)=\langle\sum_n \epsilon_nn_n(t)\rangle,
\end{equation}
and the mean squared deviation 
\begin{equation}\label{M4}
\sigma^2(t)=\langle \sum_n
\epsilon_n^2n_n(t)\rangle-\langle\epsilon\rangle^2(t)
\end{equation}
of a particle packet in energy space. Here the bracket symbolizes the
configuration average. Both quantities can be measured in time resolved 
photoluminescence experiments.
In order to calculate these quantities we first
assume that at time $t=0$ there is a particle packet, characterized by
$\{n_m(t=0)\}$, and then use the function $P_{m'm}$ (see
Ref. [\onlinecite{Boettger}]), 
the Green
function of Eq.(\ref{M1}), to calculate the electron distribution for 
$t>0$ according to the equation
\begin{equation}\label{M5}
n_m(t)=\sum_n n_n(t=0)P_{nm}(t).
\end{equation}
%
%%%%%%%%%%%%%%%%%%%%%%%%%%%%%%%%%%%%%%%%%%%%%%%%%%%%%%%%%%%%%%%%%%%%%%%
\section{The energy distribution function}
In order to derive an equation for the energy distribution function
we use the relationship
\begin{equation}
F(\epsilon',\epsilon|t)=\int d{\bm R} F({\bm R}|\epsilon',\epsilon,t)
\end{equation}
between the energy distribution function $F(\epsilon',\epsilon|t)$ and
the diffusion propagator $F({\bm R}|\epsilon',\epsilon,t)$. 
The latter is calculated from the configuration average
of the Green function $P_{nm}$. To calculate the average
we first introduce continuous coordinates according to the definition
\begin{equation}\label{DP1}
P(V';V)=
\sum_{nm}
\delta(V'-V_{n})
P_{nm}\delta(V-V_{m}),
\end{equation}
where $V=({\bm R}, \epsilon)$ and $V_m=({\bm R}_m,\epsilon_m)$.
Thereafter, we apply the effective medium-approximation by Gouchanour et. 
al.\cite{Gouchanour}. Although this 
technique has been developed originally for 
investigations of hopping systems with topological disorder it can be checked
that the extension of this method to systems with
both topological and energetic disorder only amounts to a change of
notation. Accordingly, we can directly use the results of
Ref.[\onlinecite{Gouchanour}], as discussed also in Ref.[\onlinecite{BBB1}].

The effective medium approximation of Ref.[\onlinecite{Gouchanour}] reduces
the calculation of the diffusion propagator to the solution to the
system of integral equations
\begin{equation}\label{DP2}
\langle P(\bm{R},\epsilon'|\bm{0},\epsilon)\rangle=N(\epsilon')
F(\bm{R}|\epsilon',\epsilon),
\end{equation}
\begin{widetext}
\begin{eqnarray}\label{DP3}
sF(\bm{R}|\epsilon',\epsilon)=\delta(\bm{R'}-\bm{R})
\delta(\epsilon'-\epsilon)
+\int d\rho_1&[&F(|\bm{R-R_1}||\epsilon',\epsilon_1){\tilde
W}(R_1|\epsilon_1,\epsilon)N(\epsilon)\nonumber\\
&-&F(|\bm{R}||\epsilon',\epsilon)
{\tilde W}(R_1|\epsilon',\epsilon_1)N(\epsilon_1)],
\end{eqnarray}
\end{widetext}
\begin{equation}\label{DP4}
{\tilde W}(R|\epsilon',\epsilon)=\frac{W(R|\epsilon',\epsilon)}
{1+f(\epsilon)W(R|\epsilon,\epsilon')+f(\epsilon')W(R|\epsilon',\epsilon)},
\end{equation}
and
\begin{equation}\label{DP5}
f^{-1}(\epsilon)=s+\int d\epsilon_1dR{\tilde
W}(R|\epsilon,\epsilon_1)N(\epsilon_1).
\end{equation}
Here  $s$ is the Laplace frequency, which 
corresponds to a Laplace transformation with respect to time,  
$W(R|\epsilon',\epsilon)=W_{nm}|_{R_{nm}=R, \epsilon_n=\epsilon',
\epsilon_m=\epsilon}$, and ${\tilde W}$ is the renormalized transition 
probability. $f(\epsilon)$
is the effective-medium constant, which for the present system is
strictly speaking not a constant, but a function of energy. According to these
equations the functions $F$ and ${\tilde{W}}$ depend also on $s$, but the 
dependence of $F$ on $s$ and the dependence of ${\tilde W}$ on $s$ have been 
suppressed to simplify the notation. 

Despite the effective-medium approximation the calculation of the
diffusion propagator is still a difficult task, since it requires finding 
solutions of integral equations. Due to the lack of symmetry, however, it
is impossible to solve these integral equations 
analytically if the density of states depends on energy, as it is the case
for most relevant physical systems. To simplify the equations we restrict
the consideration to the limit $\beta\omega\ll 1$ and use the
concept of quasi-elasticity. The
introduction of this concept relies on the notion that the energy scale
$\omega$ is small, so that
\begin{equation}\label{QA1}
\frac{\omega}{f(\epsilon)}\frac{df(\epsilon)}{d\epsilon}\ll1.
\end{equation}
Then the renormalized transition probability takes the form
\begin{eqnarray}\label{QA2}
{\tilde
W}(R|\epsilon',\epsilon)&=&\theta(\epsilon'-\epsilon)
\theta(\omega-\epsilon'+\epsilon){\tilde
W}(R|\epsilon')\nonumber\\
&+&\theta(\epsilon-\epsilon')
\theta(\omega-\epsilon+\epsilon'){\tilde
W}(R|\epsilon)e^{-\beta(\epsilon-\epsilon')},
\end{eqnarray}
where
\begin{equation}\label{QA3}
{\tilde W}(R|\epsilon)=\frac{W(R)}{1+2f(\epsilon)W(R)},
\end{equation}
and $W(R)=\nu\exp(-2\alpha R)$. If we furthermore restrict the consideration
to high temperatures, where $\beta\omega\ll 1$, and use the inequality
(\ref{QA1}) we find that the self-consistency equation (\ref{DP5})
simplifies considerably. For $s=0$ we find that the solution of this
equation is given by
\begin{equation}\label{QA4}
\rho_c(\epsilon,0)=\frac{2\alpha}{(\omega N(\epsilon))^{1/d}}
(\frac{d}{S_d})^{1/d},
\end{equation}
and for small $s$, that is for $s$ satisfying 
$|\rho_c(\epsilon,0)-\rho_c(\epsilon,s)|\ll\rho_c(\epsilon,0)$, we obtain
the equation
\begin{equation}\label{QA5}
(\rho_c(\epsilon,0)-\rho_c(\epsilon,s))
\exp(\rho_c(\epsilon,0)-\rho_c(\epsilon,s))=\frac{s}{\omega_0(\epsilon)}.
\end{equation}
Here
\begin{equation}\label{QA6}
\rho_c(\epsilon,s)=\ln(2f(\epsilon,s)\nu)
\end{equation}
is the dimensionless characteristic hopping length,
\begin{equation}\label{QA7}
\omega_0(\epsilon)=\frac{2d\nu}{\rho_c(\epsilon,0)}\exp(-\rho_c(\epsilon,0)),
\end{equation}
and $S_d$ is the solid angle in $d$ dimensions ($S_2=2\pi$, $S_3=4\pi$). 
The dimensionless characteristic hopping length $\rho_c(\epsilon,s)$ 
and  the characteristic hopping length $R_c(\epsilon,s)$ are connected
by the relationship $\rho_c(\epsilon,s)=2\alpha R_c(\epsilon,s)$. 

In the same approximation the equation for the energy distribution function
takes the form
\begin{widetext}
\begin{eqnarray}\label{QA10}
sF(\epsilon',\epsilon)=\delta(\epsilon'-\epsilon)
+\omega\frac{d}{d\epsilon}[N(\epsilon)v(\epsilon,s)
(\frac{d}{d\epsilon}\frac{F(\epsilon',\epsilon)}{N(\epsilon)}+\beta
\frac{F(\epsilon',\epsilon)}{N(\epsilon)})].
\end{eqnarray}
\end{widetext}
Here
\begin{equation}\label{QA11}
v(\epsilon,s)=\frac{\omega^2}{3}N(\epsilon)\int d{\bm R}{\tilde W}(R|\epsilon)
\end{equation}
is the spectral energy relaxation rate.   
Note that, due to the Eqs.(\ref{DP4})
and (\ref{DP5}) the effective transition probability depends on the Laplace
frequency $s$. Therefore, the energy relaxation rate  is also frequency 
dependent,
so that we can expect that dispersion of the energy relaxation rate 
manifests itself also in the moments of the energy distribution function.

In order to find a concrete expression for the spectral
energy relaxation rate we use Eq.(\ref{QA2}) and restrict the
investigation to low frequencies. Doing so,  we obtain
\begin{equation}\label{QA14}
v(\epsilon,s)=\frac{\omega}{3}\nu
\exp(-\rho_c(\epsilon,s)).
\end{equation}
The Eqs.(\ref{QA4}),(\ref{QA5}), (\ref{QA10}) and (\ref{QA14}) yield a closed
set of equations for the calculation of the energy distribution function,
which can be used for the investigation of energy relaxation processes in an
arbitrary density of states, provided the inequalities (\ref{QA1}) and
$\beta\omega\ll 1$ are satisfied. Below we apply this set of equations
to the investigation of energy relaxation processes in model densities of 
states.
%%%%%%%%%%%%%%%%%%%%%%%%%%%%%%%%%%%%%%%%%%%%%%%%%%%%%%%%%%%%%%%%%%%%%%%%%
\section{Energy relaxation in a constant density of states}
In this section we focus on energy relaxation processes in an unbounded 
constant density of states with $N\omega$ sites per volume. 
Investigations of energy relaxation processes in such densities of states
should be relevant for charge carriers in
certain types of Anderson-insulators above the glass temperature, as, e.g., 
for those used in the Refs.[\onlinecite{Ovadyahu1}] and
[\onlinecite{Ovadyahu2}], for which field effect measurements indicate
that the density of states varies only weakly with energy. 

Since for a constant density of states the transport coefficients are 
independent of energy the calculation simplifies considerably. In this
case the differential equation (\ref{QA10}) can be solved exactly. Its 
solution takes the form
\begin{widetext}
\begin{equation}\label{CD1}
F(\epsilon_0,\epsilon;s)=\frac{1}{2\omega v(s)}
\frac{\exp(-\frac{\beta}{2}(\epsilon-\epsilon_0)-
\sqrt{\frac{\beta^2}{4}+\frac{s}{\omega v(s)}}|\epsilon_0-\epsilon|)}
{
\sqrt{\frac{\beta^2}{4}+\frac{s}{\omega v(s)}}}.
\end{equation}
\end{widetext}
To find the dependence of the distribution function on time we have
to perform an inverse Laplace transformation, according to
\begin{equation}\label{CD2}
F(\epsilon_0,\epsilon;t)=\frac{1}{2\pi i}\int_c ds
e^{st}F(\epsilon_0,\epsilon;s).
\end{equation}
This integral yields the probability to find a particle on a site with site
energy $\epsilon$ at time $t$, if it was on a site with site energy
$\epsilon_0$ at time $t=0$. In calculating this integral we restrict the 
consideration to large times. For large times we expect that the particle 
packet has already moved down in energy space. It has somewhere in energy 
space a maximum and some width. The position of the maximum is characterized
by the mean energy $\langle\epsilon\rangle(t)$ and the width by the
dispersion $\sigma^2(t)$. We assume that $t$ is so large, that 
$\epsilon_0>\langle\epsilon\rangle(t)-\sigma(t)$.  In this case we can use 
the saddle point approximation. The saddle point satisfies the equation
\begin{equation}\label{SP1}
t=\frac{d}{ds}(\epsilon_0-\epsilon)(\frac{\beta}{2}-
\sqrt{\frac{\beta^2}{4}+\frac{s}{\omega v(s)}}).
\end{equation}
The solution of this equation is a function $s_0(\epsilon_0,\epsilon, t)$.
If we assume that the integral is determined by the saddle point we deduce
that
\begin{equation}\label{SP2}
F(\epsilon_0,\epsilon,t)\propto\exp((\frac{\beta}{2}-
\sqrt{\frac{\beta^2}{4}+\frac{s_0(\epsilon_0,\epsilon,t)}
{\omega v(s_0(\epsilon_0,\epsilon,t))}})(\epsilon_0-\epsilon)).
\end{equation}
Accordingly, the maximum of the particle packet satisfies the equation
\begin{eqnarray}\label{SP3}
0&=&\frac{d}{d\epsilon}
[(\frac{\beta}{2}+
\sqrt{\frac{\beta^2}{4}-\frac{s_0(\epsilon_0,\epsilon,t)}
{\omega v(s_0(\epsilon_0,\epsilon,t))}})(\epsilon_0-\epsilon)]\nonumber\\
&=&-\frac{\beta}{2}+
\sqrt{\frac{\beta^2}{4}+\frac{s_0(\epsilon_0,\epsilon,t)}
{\omega v(s_0(\epsilon_0,\epsilon,t))}}.
\end{eqnarray}
Here we have used the fact that $s_0$ satisfies the Eq.(\ref{SP1}).
Thus, we know that the characteristic $s_0$ in the vicinity of the maximum
is close to zero. Accordingly, in order to calculate the integral we can
restrict the consideration to small $s$, and use this knowledge to expand
the root in the exponent of Eq.(\ref{CD1}). Doing so, we obtain
\begin{eqnarray}\label{SP4}
\sqrt{\frac{\beta^2}{4}+\frac{s}{\omega v(s)}}\approx
\frac{\beta}{2}(1&+&\frac{2s}{\beta^2\omega v(0)}-\frac{2s^2}{\beta^2\omega
v(0)\omega_0}\nonumber\\
&-&\frac{2s^2}{(\beta^2\omega)^2v^2(0)}).
\end{eqnarray}
Here it is important to realize that the third and fourth term of this
expansion is of different physical origin. Both terms, as can be seen
further below, contribute to the mean squared deviation. However, while the
third term results entirely from  dispersion the fourth term results
entirely from thermal energy diffusion. Thus the relationship between these
two terms determines, whether thermal activation  affects the shape of the
energy distribution function, or not.
If we use the expansion (\ref{SP4}) the integrals become Gaussian. Doing 
the integrals we find that
\begin{equation}\label{SP5}
F(\epsilon_0,\epsilon;t)=\frac{1}{\sqrt{2\pi \sigma^2(t)}}
\exp(-\frac{(\epsilon_0-\epsilon-\beta\omega v(0) t)^2}{2\sigma^2(t)})
\end{equation}
where
\begin{equation}\label{SP6}
\sigma^2(t)=2\omega v(0)t(1+\frac{\beta^2\omega v(0)}{\omega_0}).
\end{equation}
Accordingly, at large times the particle packet moves with constant
velocity. An increase of the temperature reflects in a decrease of
the energy relaxation rate. The shape of the packet is Gaussian, and  
its mean squared deviation is given by Eq.(\ref{SP6}). In this equation
the first term in the bracket on the right hand side results entirely
from thermal diffusion and the second entirely from dispersion. 
Remarkably, the thermal contribution to the packets width is independent
of temperature. Since
\[\frac{\beta^2\omega v(0)}{\omega_0}=\frac{(\beta\omega)^2\rho_c}{6d}\]
its contribution is only important if
the temperature is large, that is, if $T\gg T_{c}$, where 
$kT_c=\omega \sqrt{\rho_c/(6d)}$. Accordingly, the 
second derivative in Eq.(\ref{QA10}) is negligible for 
$T\ll T_{c}$. Consequently, the 
relaxation is determined by the first order differential equation
\begin{equation}\label{SP7}
sF(\epsilon_0,\epsilon)=\delta(\epsilon_0-\epsilon)+
\frac{d}{d\epsilon}(F(\epsilon_0,\epsilon)\omega\beta v(s)),
\end{equation}
for $T\ll T_{c}$.
If the relaxation at zero temperature is also described percolation like
then this equation differs from the equation for the calculation of the
energy distribution function at zero temperature 
(see Ref.[\onlinecite{BBB2}]) only in that $v(0)$ is replaced by 
$\omega\beta v(0)$, so that all of the results derived in 
Ref.[\onlinecite{BBB2}] are also valid in this regime, after a trivial
change of parameters. Since $T_c\propto\sqrt{\rho_c}$ the range of
applicability of these results increases with increasing disorder.
%%%%%%%%%%%%%%%%%%%%%%%%%%%%%%%%%%%%%%%%%%%%%%%%%%%%%%%%%%%%%%%%%%%%%%
\section{Energy relaxation in an exponential density of states}
In this section we investigate the relaxation in an exponential density of
states of the type
\begin{equation}\label{ER1}
N(\epsilon)=N_0\exp(\frac{d\epsilon}{\Delta}).
\end{equation}
Here $N_0$ and $\Delta$ are parameters characterizing the density of states.
The consideration of the relaxation in such a density of states is relevant
for investigations on relaxation phenomena in band tails of several 
amorphous semiconductors, as e.g., for the conduction band tail in amorphous
hydrogenated silicon (see, e.g., Ref. [\onlinecite{Monroe}]). 

Since the strong dependence of the density of states on energy manifests 
itself also in the transport coefficients it turns out to be hard to find 
an analytical solution for the differential equation (\ref{QA10}) in this
case. Therefore, we focus at first on the solution for a particle
which moves in a region with
\begin{equation}\label{ER2}
kT\frac{d\rho_c(\epsilon,0)}{d\epsilon}\ll 1.
\end{equation}
That is, we first restrict the consideration to the shallow states in the
tail.
The lower limiting energy of this region agrees with the 
transport energy, as defined in the
Refs.[\onlinecite{Monroe}], [\onlinecite{Shapiro}] and 
[\onlinecite{Thomas}]. In the literature
this region is also called the hopping down regime \cite{Monroe}.

Within the range of the inequality (\ref{ER2}) we can use the WKB-
approximation (see the details of the
calculation in appendix \ref{CE}). Doing so, we find that 
the energy distribution function takes the form 
\begin{equation}\label{ER12}
F(\epsilon_0,\epsilon;t)=\frac{1}{\sqrt{2\pi\sigma^2(t)}}
\exp(-\frac{(\epsilon-\langle\epsilon\rangle(t))^2}{2\sigma^2(t)})
\end{equation}
at large times, 
where the mean energy is given by
\begin{equation}\label{ER11}
\langle\epsilon\rangle(t)\approx-\Delta\ln(\frac{1}{\rho_0}
\ln(\frac{\beta\omega^2}{3\Delta}\nu t)),
\end{equation}
(here $\rho_0=\rho_c(\epsilon)\exp(\epsilon/\Delta)$)
and the mean squared deviation by
\begin{equation}\label{ER13}
\sigma^2(t)=\sigma^2_{therm}(t)+\sigma^2_{disp}(t).
\end{equation}
Here
\begin{equation}\label{ER16}
\sigma^2_{therm}(t)\approx \frac{kT\Delta}
{\rho_c(\langle\epsilon\rangle(t))},
\end{equation}
and
\begin{equation}\label{ER17}
\sigma^2_{disp}\approx
\frac{\Delta\beta\omega^2}{6d}.
\end{equation}
Eq.(\ref{ER12}) differs from the predictions of the energy loss hopping model,
which have been derived with the assumption that the relaxation at zero
temperature is percolation like, in two points. First, while in the 
energy loss hopping model the energy relaxation 
rate is independent of temperature \cite{Monroe,BBB2} the energy relaxation 
rate is determined by the parameter $\beta\omega^2\nu t/\Delta$
at finite temperatures in the limit $\beta\omega\ll 1$. 
Second, while the 
mean squared
deviation of the particle packet in the energy loss hopping model is time
independent, the dispersion becomes time dependent at finite temperatures.
The time dependence results from the thermal
contribution to the mean squared deviation. This contribution to the
mean squared deviation is 
dominant if
\begin{equation}\label{ER18}
kT\gg
\omega\sqrt{\rho_c(\langle\epsilon\rangle(t))/(6d)},
\end{equation}
that is for high temperatures. In this temperature range the width
of the energy
distribution function increases with temperature. However, since
the particle is sinking down its hopping length 
increases. From Eq.(\ref{ER16}) we deduce that this increase tends to
suppress the thermal contribution to the packet's width, 
and thus renders the disperse contribution more important. In contrast to 
the thermal contribution the disperse contribution is 
time independent, so that the packet moves in energy space without further
distortion, once the thermal spreading of the packet has become
insignificant. In this limit the motion of the packet in energy space 
is soliton like. Here an 
increase of the temperature 
results in a decrease of the packet width.

The fact that the thermal spreading of the particle packet becomes
unimportant outside the validity range of the inequality (\ref{ER18})
indicates that in this region the second derivative with respect to the 
energy in Eq.(\ref{QA10}) is negligible, so that instead of Eq.
(\ref{QA10}) the simple equation
\begin{equation}\label{ER19}
sF(\epsilon_0,\epsilon)=\delta(\epsilon-\epsilon_0)+
\frac{d}{d\epsilon}(F(\epsilon_0,\epsilon)\beta\omega v(\epsilon,s)).
\end{equation}
can be used. 

In the derivation of our results we have used the inequality (\ref{ER2}),
which controls the validity of the WKB-approximation. However, the
results of this calculation show that thermal spreading becomes 
unimportant when time goes by, that is when the hopping length becomes
large. From our point of view this is intuitively clear, since 
the impact of the dispersion of the transport coefficients in the hopping
regime increases strongly with disorder. This fact suggests, that 
Eq.(\ref{ER19}) also holds outside the validity range of  the inequality 
(\ref{ER2}), so that Eq.(\ref{ER19}) can probably also be used for deep 
states in the tail, if only the condition (\ref{QA1}), which governs the 
range of applicability
of the quasi-elastic approximation, is met.
%%%%%%%%%%%%%%%%%%%%%%%%%%%%%%%%%%%%%%%%%%%%%%%%%%%%%%%%%%%%%%%%%%%%%%%%
\section{Conclusions}
Our investigations show that the structure of the energy distribution function
in the hopping regime is determined by the dispersion of the energy
relaxation rate in a large temperature range. The characteristic temperature, 
governing the impact of the dispersion on the structure of the energy 
distribution function, depends on the strength of the disorder, and
therefore on the details of the density of states.
In both of the cases studied the impact of energy diffusion processes
on the structure of the energy distribution function were only significant 
for $kT\gg\omega\sqrt{\rho_c}(\epsilon,0)$. This sets the situation in the 
hopping
regime apart from that on the extended side of the metal-insulator
transition, in which the structure of the energy distribution function is
entirely determined by energy diffusion\cite{Landau,Kampen}.

In a constant density of states the parameter $\rho_c(\epsilon,0)$ is
independent of $\epsilon$. Therefore, a global characteristic temperature
exists, which governs the influence of the dispersion. Although in this 
case the mean squared deviation increases linearly with time for large
times the difference between the energy diffusion driven 
structuring of the energy diffusion function
and a structuring due to the dispersion of the transport coefficients
manifest themselves in the temperature dependence of the 
second moment of the energy distribution function. The latter 
decreases quadratically with increasing temperature at large temperatures.

In an exponential density of states the situation is more intricate. In this 
case there is no globally defined characteristic temperature, since the 
criterion $kT\gg\omega\sqrt{\rho_c}(\epsilon,0)$ depends on the 
instantaneous position of the packet in the tail. Since
the particles are sinking down their dimensionless hopping length 
$\rho_c(\epsilon,0)$ increases when time elapses. Accordingly, 
the thermal spreading of the energy distribution function, which is 
important in the initial phase of the relaxation in the shallow states of 
the tail, is getting unimportant when time goes by. Deep
in the tail the structure of the energy distribution function
is entirely determined by the dispersion of the energy relaxation rate.
Energy diffusion processes are negligible in this case. In this limit
the dispersive nature of relaxation process leads to a soliton
like motion of the particle packet in energy, i.e., the particle packet moves
in energy space without distortion.  

We would like to mention that in both  cases studied an increase
of the temperature results in decrease of the energy relaxation rate.
This result is in line with the results the numerical simulations in 
Ref.[\onlinecite{Haba}] for a constant density of states, which have
been performed in the same limit. It sets the situation in the limit
of small energy transfer apart from that in conventional variable-range 
hopping systems, in which the relaxation rate 
usually speeds up with increasing temperature \cite{Monroe}.
%%%%%%%%%%%%%%%%%%%%%%%%%%%%%%%%%%%%%%%%%%%%%%%%%%%%%%%%%%%%%%%%%%%%%%%%%
\begin{appendix}
\section{Calculation of the inverse Laplace transform for the exponential
density of states}\label{CE}
Within the range of the inequality (\ref{ER2}) we can use the
WKB-approximation. However, in applying the WKB-approximation the
following difficulty appears. To lowest order, that is within exponential
accuracy, the WKB-approximation immediately predicts
\begin{equation}\label{ER3}
F(\epsilon_0,\epsilon)\propto\exp(-\frac{\beta}{2}(\epsilon-\epsilon_0)
-|\int_{\epsilon}^{\epsilon_0} d\epsilon_1
\sqrt{\frac{\beta^2}{4}+\frac{s}{\omega v(\epsilon_1,s)}}|).
\end{equation}
If we use the inequality (\ref{ER2}) we can generate a systematic 
expansion of the preexponential factor of the diffusion propagator 
with respect to the parameter
$kT\rho_c'(\epsilon)$, where the prime indicates the derivative with respect
to energy.
However, since the error estimate in the WKB approximation is of
multiplicative type, non of the approximations can satisfy probability
conservation exactly. Therefore, we only use the exponent, and determine
the preexponential factor by normalization.

In investigating the energy distribution function further we again restrict
the consideration to large times, where 
$\epsilon_0\gg\langle\epsilon\rangle(t)-\sigma(t)$.
In this case the calculation of the time dependence of the energy
distribution function amounts to the calculation of the integral
\begin{equation}\label{ER4}
I(t)=\frac{1}{2\pi i}\int_c ds
\exp(st-\int_{\epsilon}^{\epsilon_0}d\epsilon_1
(\sqrt{\frac{\beta^2}{4}+\frac{s}{\omega v(\epsilon_1,s)}}-
\frac{\beta}{2})).
\end{equation}
In order to calculate this integral we use the same method as for the
constant density of states. Doing so, we find that
\begin{equation}\label{ER5}
I(t)=\frac{1}{\sqrt{2\pi D^2(\epsilon)}}
\exp(-(\frac{(t-t_{\epsilon})^2}
{2D^2(\epsilon)})).
\end{equation}
Here
\begin{equation}\label{ER6}
t_{\epsilon}=\int^{\epsilon_0}_{\epsilon}d\epsilon_1
\frac{d\epsilon_1}{\beta\omega v(\epsilon_1,0)},
\end{equation}
and
\begin{equation}\label{ER7}
D^2(\epsilon)=D^2_{therm}(\epsilon)+D^2_{disp}(\epsilon),
\end{equation}
where the thermal contribution is given by
\begin{equation}\label{ER8}
D^2_{therm}(\epsilon)=2\int^{\epsilon_0}_{\epsilon} 
\frac{d\epsilon_1}{\beta^3\omega^2v^2(\epsilon_1,0)},
\end{equation}
and the disperse contribution by
\begin{equation}\label{ER9}
D^2_{disp}(\epsilon)=2
\int^{\epsilon_0}_{\epsilon} 
\frac{d\epsilon_1}{\beta\omega v(\epsilon_1,0)\omega_0(\epsilon_1)}.
\end{equation}
Now we use the fact that our calculation of the inverse Laplace
transformation is only correct in the vicinity of the peak. Since
the peak is situated at $t=t_{\epsilon}$ the maximum of the packet in
energy space $\langle\epsilon\rangle(t)$ is situated at $\epsilon=
\langle\epsilon\rangle(t)$, where 
\begin{equation}\label{ER10}
t=t_{\langle\epsilon\rangle(t)}.
\end{equation}
If we use Eq.(\ref{ER10}) to calculate the mean energy at large times we
obtain Eq.(\ref{ER11}). Since
our calculation of the inverse Laplace transformation is only correct in the
vicinity of $\langle\epsilon\rangle(t)$ we now replace $\epsilon$
by $\langle\epsilon\rangle(t)$ in Eq.(\ref{ER5}) and then adjust the
preexponential factor, as discussed above. Doing so, we obtain
Eq.(\ref{ER12}), where
\begin{eqnarray}\label{ER14}
\sigma^2_{therm}(t)&=&2D_{therm}(\langle\epsilon\rangle(t))\beta^2\omega^2
v^2(\langle\epsilon\rangle(t),0)\nonumber\\
&\approx& \frac{kT\Delta}
{\rho_c(\langle\epsilon\rangle(t))},
\end{eqnarray}
and
\begin{equation}\label{ER15}
\sigma^2_{disp}=2D_{disp}(\langle\epsilon\rangle(t))\beta^2\omega^2
v^2(\langle\epsilon\rangle(t),0)\approx
\frac{\Delta\beta\omega^2}{6d}.
\end{equation}
\end{appendix}
%%%%%%%%%%%%%%%%%%%%%%%%%%%%%%%%%%%%%%%%%%%%%%%%%%%%%%%%%%%%%%%%%%%%%%%%

\end{document}